\documentclass{article}

\usepackage{graphicx} 
\usepackage{comment}

\begin{document}

\title{Workshop on the limiting compactness objects: Black holes and Buchdahl stars\\ \vskip 1em
\large IUCAA, Oct 30 - Nov 3, 2023 \vskip 2em}

\author{Dawood Kothawala$^1$ and Sahil Saini$^2$}

\date{	
    $^1$Department of Physics, Indian Institute of Technology Madras, Chennai 600036, India \\ \textit{dawood@iitm.ac.in}\\%
	$^2$Department of Physics, Guru Jambheshwar University of Science \& Technology, Hisar, Haryana 125001, India \\ \textit{sahilsaiini@gjust.org}\\[2ex]%
}


\maketitle

The workshop was organized at IUCAA on Oct 30 – Nov 3 as a compact discussion/discourse meeting with a threadbare exposition and discussion of the various aspects and the questions arising. It was occasioned by the visit of Professor H\r{a}kan Andr\'{e}asson of the Gothenburg Technical University, Sweden. He has been exploring with his collaborators the Einstein – Vlasov system for over a decade and a half as a possible matter source for compact objects. This system characterizes itself by free particles in motion and interacting only through gravity. For a limiting compactness, this may be the most appropriate state. 

The main thrust of the workshop was to understand this new object, BS, of limiting compactness without a horizon. It is almost as compact as a BH and yet has no horizon and hence is open for interaction with the outside world. Ever since the proposal of the membrane paradigm envisaging a timelike fiducial surface near BH horizon, BS offers an excellent possibility of the existence of such a real astrophysical object. It could very well compete with BH as a mimicker for various physical and astrophysical phenomena. Thus, it opens up a new vista of study and investigation of all the questions that one asks for BH, for this new creature, BS. The workshop was intended to identify certain interesting questions as well as the people interested in studying them. On this count, the workshop has been a huge success as several interesting questions have been identified, a few groups have been formed to take up different problems, and the work has already started. Nothing more could one have asked from such an exercise. A brief summary of some of the talks follows now followed by a brief discussion of the projects identified as a result of the discussions during the workshop.

{\bf Naresh Dadhich: The most compact object}

Black hole: The most compact object is a black hole which is characterized geometrically by the timelike Killing vector turning null, defining the event horizon. This has no reference to the internal structure which is however completely unknown. Another common-sense definition is when a freely falling particle attains the velocity of light, i.e., $v^2=2\Phi(R)=1$ where $\Phi(R)$ is the gravitational potential felt by the radially falling particle \cite{Naresh}. That is, $\Phi(R)=1/2$ defines a black hole with a horizon as the most compact object. 

Alternatively, BH can also be defined when gravitational and non-gravitational energy (mass) become equal.  That is the equipartition of the total energy into gravitational and non-gravitational energy. For spherically symmetric static spacetime, gravitational energy could be computed from the unique exterior spacetime metric by employing the Brown-York quasi-local energy prescription, and non-gravitational energy is mass $M$ for neutral and $(M-Q^2/2R)$ for the charged object. This is an insightful characterization in terms of the balance between gravitational and non-gravitational energy.  

Buchdahl star: BH is the most compact object with a horizon, so the question arises what is the most compact non-BH object without a horizon? Long back Buchdahl had obtained the compactness limit, $M/R \leq 4/9$  by demanding the density of the isotropic fluid being non-increasing outwards from the centre, and the interior is matched to the vacuum exterior at the boundary defined by vanishing of pressure. In terms of the potential, this will translate to $\Phi(R) \leq 4/9$.

We define the Buchdahl star (BS) by the equality in the bound; i.e., $\Phi(R)=4/9$, equivalently $v^2=8/9$. Thus, BH and BS are respectively defined by $\Phi(R)=1/2,4/9$.

Like BH, could BS be also defined by the balance between gravitational and non-gravitational energy? Indeed, it is so when the former is half of the latter; i.e., $E_\mathrm{g}=\frac{1}{2} E_\mathrm{ng} $. This reminds of the Virial theorem if gravitational energy is taken as the analogue of kinetic and non-gravitational energy that of potential energy. The theorem is however applicable only when the interaction between the constituents is only through gravity. The star’s interior is generally supposed to be fluid distribution which involves other interactions as well. 

A common-sense compactness limit should be offered by the constant density incompressible fluid distribution which is uniquely described by Schwarzschild’s interior solution, and then the Buchdahl compactness bound is given by pressure at the centre being finite, and the equality (BS) is attained when the pressure at the centre diverges. That means as a fluid distribution BS would have to have diverging pressure at the centre. That is not physically acceptable. What it means is that fluid even with the stiffest equation of state of constant density cannot account for the BS interior without pressure diverging at the centre. Then the only option that remains is that the equilibrium of BS is governed by the generalized Virial theorem (gravitational energy being half of the non-gravitational energy) where free particles in motion interact only through gravity. This is the Vlasov kinetic matter. The BS interior could therefore be only of the Vlasov kinetic matter which has extensively been studied by H\r{a}kan Andr\'{e}asson for over a decade and a half. 

That was how the workshop was occasioned on H\r{a}kan’s visit to IUCAA for a week. It was attended by a small group of active researchers that included, Dawood Kothawala, Sahil Saini, Rituparno Goswami (online), Ranjan Sharma, B C Paul, Shasvath Kapadia, Sanjit Mitra, Debarati Chatterjee, Nishant Singh, Arman Turusonov (online), Naresh Dadhich and others.

\textbf{ H{\aa}kan Andr\'easson: On the existence and structure of stationary solutions of the Einstein-Vlasov system: Part I}

I introduced the Einstein-Vlasov (EV) system and  discussed how to construct static solutions in the spherically symmetric case. I presented an analytic result on the maximum compactness of such solutions \cite{Haken1}. The statement is that $2M/R<8/9$ for any solution for which $p_r+2p_T\leq\rho$, where $p_r, p_T$ and $\rho$ are the radial pressure, the tangential pressure and the energy density respectively.  In addition to this condition it is required that $p_r\geq 0$. The bound $2M/R\leq 8/9$ agrees with the well-known bound obtained by Buchdahl in 1959.  However, these bounds rely on very different assumptions. The latter bound requires that $\rho$ is non-increasing outwards and that the pressure is isotropic. The saturating solutions in the two cases are completely different; the constant energy density solution saturates the bound by Buchdahl whereas an infinitely thin shell solution satisfies the bound I have derived. I also presented a generalization of the compactness bound to the charged case \cite{Haken2}. Finally, I discussed the existence of massless static solutions of the EV system (joint work with Fajman and Thaller) and their relation to Wheeler's \textit{geons}, and I presented a result on massless solutions surrounding a Schwarzschild black hole.

\textbf{Part II }

In this talk I presented results from a numerical study on the structure of axisymmetric stationary solutions of the EV system (joint work with Ames and Logg). 
In particular, highly compact solutions admitting ergoregions were discussed. As the solutions get more relativistic two different limits were found. Either the solution approaches an extremal Kerr black hole or the solution resembles a cosmic string. I also presented recent results on the comparison between spherically symmetric static solutions of the EV system and the Einstein-Dirac (ED) system (joint work with Blomqvist). In 1999 Finster et al. found for the first-time static solutions to the ED system in the case of two fermions with opposite spins. Recently this study has been extended to a larger number of particles by Leith et al. In particular, they construct highly relativistic solutions. The structure of the solutions to the ED system for a handful of particles turns out to be strikingly similar to the structure of highly relativistic solutions of the EV system. Since solutions of the EV system are classical whereas solutions of the ED system have a quantum signature, this study provides insights about the transition from quantum to classical behaviour for a quite small number of particles. 
 
\textbf{Rituparno Goswami: Semitetrad formalisms: Some insights into Buchdahl Limit and Black Holes}

In this talk we used the geometrical $1+1+2$ formalism, to gain some deeper insights into compact stars and black holes. First and foremost, we geometrically established yet another correspondence between Newtonian mechanics and general relativity by connecting the Buchdahl limit and the Virial theorem. Buchdahl stars are defined by the saturation of the Buchdahl bound, $\Phi(R) \leq 4/9$ where $\Phi(R)$ is the gravitational potential felt by a radially falling particle. It has been recently argued that the equilibrium of a Buchdahl star may be governed by the Virial theorem. In this talk we provided a purely geometric version of this theorem and thereby of the Buchdahl star characterisation. We know that in Newtonian mechanics, the Virial theorem states that for a system of particles acting only through gravity in equiibrium, the total average kinetic energy of the particles of the system $⟨T⟩$,  is half of the total average potential energy $⟨V⟩$. By describing the dynamics of a star at the Buchdahl limit in terms of the geometric quantities related to the timelike matter flow congruence and the preferred spacelike congruence of the spherical symmetry, and the thermodynamic quantities of matter, we could easily identify that the geometric notion of the energy due to motion is determined by the volume expansion and shear of the matter flow lines as well as the spatial volume expansion of the preferred spatial congruence. On the other hand, the geometric counterpart of the total potential energy of the system is generated by the matter energy density and pressure anisotropy as well as the free gravity described by the Weyl scalar. We asserted the proposition that the geometric Virial stability is the limiting case of static, spherically symmetric and thermodynamically stable matter configurations. We further asked a very important question that whether Is it possible for a collapsing compact star to continue in the state of Virial equilibrium?. We transparently showed that it is possible provided the collapsing star continuously radiates in the external Vaidya region via the internal heat flux and  thus remaining in the Virial equilibrium state. In this situation the star radiates away completely, escaping getting trapped into a black hole.

In the second part of the talk, we geometrically established another important result: If we  postulate that any matter field accreting onto the black hole horizon must have a phase transition from Type I to Type II, then degeneracy of globally determined null event horizon and locally defined marginally outer trapped surface for a Schwarzschild black hole is never broken, and hence it can continue to be the site of emergence of the Hawking radiation even when black hole is accreting. This phase transition comes at a cost that heat flux has to be radiated out from the collapsing cloud which manifests in the exterior as the Vaidya radiation. Thus, for an accreting black hole, quantum Hawking radiation from the horizon must accompany classical Vaidya radiation from the boundary of the accretion zone.

\textbf{Dawood Kothawala: Quantum probes of spacetime curvature and implications for quantum spacetimes}

Most of our observations that characterize space and time are expressed
in terms of non-local, bi-tensorial objects such as geodesic or
null intervals between events and two-point (Green’s) functions. In
this talk, I highlighted the importance of characterizing spacetime
geometry in terms of two fundamental bi-tensors, Synge’s World
function and the van Vleck determinant. In the first part, two specific
recent results that employ these objects were discussed: (i)
Non-perturbative tidal corrections to results such as Unruh effect
and ageing of twins, and (ii) Formulating generalised uncertainty
principle on curved space(time)s. A formalism is then described to
construct an effective quantum metric - qmetric - in terms of these
bi-tensors, which can describe spacetime with a zero point length.
Some non-trivialities of the qmetric, its classical limit, and its role in quantum gravity were highlighted. The sections below give a broad
overview of the key points and results that constituted the talk.
\\
\textbf{Spacetime from two-point correlators:} A generic implication of incorporating gravitational effects in the
analysis of quantum measurements is the existence of a zero-point
length of spacetime. This requires an inherently non-local description
of spacetime, beyond the usual one based on metric $g_{ab}(x)$ etc.
One of the major focus of the talk was to put forward the right tools to characterise spacetime
in terms of inherently non-local, and observationally relevant, quantities. The quantum spacetime should instead be reconstructed
from non-local bi-tensors of the form $G_{ab...i'j'...}(x, x')$. A deeper look then reveals a subtle interplay between non-locality and the limit $l^2_0 = G\hbar/c^3 \rightarrow 0$. In particular, the so called emergent gravity paradigm – in which gravitational dynamics/action/spacetime are emergent and characterised by an \textit{entropy functional} – arises as the \textit{Cheshire} grin of a fundamentally non-local quantum spacetime. We have constructed a non-local quantum metric, ''qmetric", which can be used as the effective object that can be used in place of the metric to obtain insights into the small scale structure of spacetime. 
\\
\textbf{Euclidean domain of Lorentzian spacetimes:}
In a related part of the talk, I discussed a new framework for Euclidean gravity, which might provide deep insights into
the small scale structure of spacetime as well as generic quantum nature of gravity. 
\\
\textbf{Entanglement and decoherence in curved spacetime:} The final part of the talk described some results that applies above tools to study the role of Riemann tensor in entanglement and decoherence.

\textbf{Ranjan Sharma: Compactness limit for relativistic stars in Einstein and extended gravity models}

The maximum permissible compactness bound for a relativistic self-gravitating object is given in terms of the Buchdahl bound, which tells us that no uniform density stars with radii smaller than $9M/4$ can exist \cite{buchdahl1959general}. As the critical compactness plays a crucial role in constraining the EOS of a star, many investigators have employed different techniques and used various stellar models to analyze the impacts of model parameters and subsequently determined the maximum permissible compactness bound. In this context, of particular interest, is a Buchdahl star which is a very highly dense compact object and still can manage to remain in equilibrium without forming an event horizon.

This article presents some of the recent developments in the estimation of maximum compactness bound for ultra-compact objects. In 1939, Tolman \cite{Tolman} developed a technique for obtaining solutions to Einstein field equations for a static spherically symmetric fluid source, which produced eight exact analytic solutions, some of which were known or rediscovered in different forms. Of the eight solutions, the Tolman VII was found to possess features of realistic stars. The solution exhibits a quadratic fall-off behaviour of the density profile which is similar to a neutron star model. Generalized \cite{Raghoonundun} and modified \cite{Jiang} versions of the solution had later been developed to accommodate a wide range of neutron star EOS.

Using the generalized version of the Tolman VII solution, we have provided an estimate of the maximum permissible compactness bound $M/R$. Our approach provides an insight into the role of model parameters on the compactness bound of a star. We have used the solution of Raghoonundun and Hobill \cite{Raghoonundun}, who had extended the Tolman VII model by considering the energy density in a more generalized form. The solution is essentially a three-parameter [$M,\rho_c, \mu$] family of solutions with $M$=mass, $\rho_c$=central density and $\mu = [0,1]$ as a free parameter representing ‘stiffness’ of the EOS of the star. In the extreme case $\mu$ = 1, one regains the original Tolman VII solution and the $\mu$ = 0 case is similar to Schwarzschild’s incompressible fluid sphere solution (or Tolman III solution). The latter readily provides the Buchdahl compactness bound $M/R < 4/9$. Making use of the generalized Tolman VII solution and applying Chandrasekhar’s variational method to analyze the stability of a spherically symmetric star against radial oscillations \cite{Chandrasekhar,Chandrasekhar2}, we have obtained an upper bound on the compactness for different values of $\mu$. By calculating the fundamental frequencies and imposing the condition for stability of the configuration, we have shown that the critical compactness bound decreases with increasing values of $\mu$. Our study shows how a departure from homogeneous matter distribution characterized by the stiffness parameter $\mu$ can play a crucial role in fixing the maximum compactness bound. Interestingly, fulfilment of the causality condition at the centre ($dp/d\rho \leq 1$) provides a more stringent compactness bound as compared to the value obtained by imposing the stability requirement condition.

Another development in this direction is to go beyond GR. Despite enormous success in predicting many tests of gravity, it is evident that GR faces many challenges at different scales. Consequently, many investigators had been tempted to modify Einstein’s gravity by modifying the Einstein-Hilbert action. The $f(R,T)$ gravity is one such example \cite{Harko2011} where the action gets extended by coupling non-minimally the matter field to the geometry; $T$ being the trace of the energy-momentum tensor. We have developed a technique to find a new upper bound on the compactness of a charged fluid sphere in $f(R,T)$ gravity. Following Harko et al \cite{Harko2011}, we have assumed that the coupling is linear i.e., $f(R,T) = R +2\chi T$ where $\chi$ is a dimensionless coupling parameter. A detailed analysis of the resultant system provides a new upper bound on compactness in the form

\begin{equation}
    \frac{M}{R} \leq \frac{\frac{32\pi}{9}\left(\frac{2}{(8\pi-\chi)}-\frac{\chi}{(8\pi-\chi)^2} \right)}{1+\frac{4\pi}{(8\pi-\chi)}\sqrt{\frac{16\alpha^2}{9}\left( \frac{3\chi}{8\pi}-2\right)+\left(\frac{\chi}{4\pi}-2\right)^2}},
\end{equation}

where $\alpha^2=Q^2/M^2$ and $Q=$ total charge within the boundary of the star. For $\chi=0$, the bound reduces to

\begin{equation}
    \frac{M}{R} \leq \frac{8/9}{1+\sqrt{1-\frac{8\alpha^2}{9}}}
\end{equation}

which is the upper bound on the compactness of a charged fluid sphere (see Giuliani and Rothman \cite{Giuliani} and Sharma et al \cite{Sharma1}. Further, in the uncharged case within the domain of Einstein gravity, one regains the Buchdahl bound $M \leq 4/9$.

\textbf{Sahil Saini: Non-singular black holes from loop quantum cosmology}

Loop quantum gravity (LQG) is a canonical quantization of general relativity when reformulated as a theory of connections. This leads to a discrete quantum spacetime where geometrical operators such as the area and volume have a discrete spectrum. Loop quantum cosmology (LQC) applies these techniques to symmetry reduced cosmological models, obtaining the quantum dynamics in terms of quantum difference equations. Detailed phenomenological implications are often worked out using a coarse-grained continuum effective geometry which incorporates the leading order quantum corrections. Strikingly, it turns out the singularity is replaced by a quantum bounce in LQC. Moreover, the effective dynamics allows to analytically show the finiteness of various physical quantities such as energy density, expansion and shear scalars etc., and the absence of strong curvature singularities (Ph.D. thesis, Sahil Saini, Louisiana State University, 2019 \cite{Saini1}).

We restrict our attention here to static non-rotating black holes (BH). The application of LQC to these BHs stems from the isometry between the Kantowski-Sachs (KS) cosmology and the interior of a Schwarzschild black hole. While quantizing, the curvature is expressed using holonomies of the connection around appropriately chosen loops. Thus, the effective equations inherit certain ‘quantum parameters’ $\delta_b$ and $\delta_c$, that refer to edge lengths of these loops, which vanish in the classical limit as the loops are shrunk to zero. These parameters are chosen such that physical predictions are independent of coordinate rescaling and of various fiducial structures introduced during quantization. All such quantizations resolve the BH singularity, replacing it with a transition surface/bounce, beyond which lies an anti-trapped region. Yet, there still remains some ambiguity in the choice of these parameters, and when applied to the BH interior, their choice turns out to be subtle. Two such choices of parameters were obtained in \cite{Boehmer} and \cite{Corichi}. In \cite{Boehmer}, the so called $\bar\mu$-prescription was derived and the parameters were chosen as appropriate phase space functions such that the physical area of the loops is equal to the minimum area eigenvalue in LQG. A detailed analytical investigation revealed \cite{Saini2} that in addition to resolving the singularity, it led to undesirable quantum corrections at the horizon which can be at very low curvatures for large mass BHs. The quantization in \cite{Corichi} did not suffer these problems. It defined the parameters as certain Dirac observables depending on the BH mass. However, the quantum bounce turns out to be highly asymmetric, giving the ADM mass of the anti-trapped region of the order of the fourth power of that of the parent BH. 
The work \cite{Saini3} modifies the quantization of \cite{Corichi} to obtain a symmetric bounce, and investigated three examples of it. In all three cases, the curvature at bounce depended on the mass of the BH. This is in contrast with the trend in LQC, where universal bounds for the curvature at the bounce have been obtained in various models. Moreover, the bounce could occur at very low curvatures for large mass BHs. Such large quantum effects at low curvatures are undesirable. This motivated the formulation of a prescription where the curvature invariants have universal bounds at the bounce. In \cite{AOS1}, the parameters are defined by considering the loops on the bounce itself. Defined this way, the curvature invariants acquire a well-defined universal upper bound at the bounce surface. Moreover, the ADM mass on both sides is approximately equal (nearly symmetric) for large mass BHs. Quantum modifications are also negligible in low curvature regions, as one would expect. The exterior region of the BH horizon is also explicitly constructed in \cite{AOS1}. However, some puzzling features remain. Though the resultant spacetime is asymptotically flat in a precise sense (suffices to define ADM mass), the approach to flatness is weaker compared to conventionally used definition of asymptotic flatness.

\section{List of interesting questions that emerged from the workshop}
The discussions following the talks led to a number of interesting questions about compact stars and their properties, which were taken up as projects following the workshop. A brief discussion of the identified questions follows now:

\subsection{Entropy of Buchdahl Stars}
H\r{a}kan Andr\'{e}asson has been studying compact spherical objects made of Einstein-Vlasov matter for over a decade using computer simulations. As already mentioned above, a Vlasov system consists of matter particles which interact only through gravity. In this sense it is similar to dust matter often considered in cosmological settings. However, unlike dust, the matter particles in the Vlasov system have some initial velocity and kinetic energy to begin with. Thus, it is a generalized version of dust. The dynamics of such a system under gravity are governed by the Einstein-Vlasov system of differential equations, which are very unwieldy and require computer simulations to study their evolution, which H\r{a}kan has been engaged in studying at the Gothenburg Technical University, Sweden.

Curiously, H\r{a}kan’s simulations suggest that matter in these compact spherical Vlasov systems tends to organize in thin shells as the compactness of the object is increased. The higher the compactness, the lower is the thickness of these shells. As the compactness become closer to the Buchdahl limit, most of the matter tends to reside in a thin shell at the surface of these objects. This raises an interesting question about the entropy of the system. It is known that since gravitational interaction is long-range (and unlike electromagnetism there is no shielding), the conventional intuition about statistical mechanics does not hold. In particular, it has been shown that the entropy of a compact spherical object becomes approximately proportional to its surface area rather than its volume as it approaches limiting compactness, even before the formation of a black hole. The major contribution to this entropy comes from the gravitational interaction which causes it to scale as the area. The non-gravitational component of the entropy can still be considered to be an extensive quantity and scales as the volume, but is dominated by the entropy of the gravitational interaction as compactness increases. However, H\r{a}kan’s simulations seem to suggest that this transition of the total entropy from volume-scaling to area-scaling might be even stronger than previously suspected, as the matter content itself also tends to reside in a thin shell at the boundary as we approach limiting compactness. Thus, it suggests that the non-gravitational extensive component of the entropy also begins to scale as the area as we approach limiting compactness.

Of course, these intuitions have to be verified through a detailed study of the entropy of these compact objects. One of the projects to come out of the workshop thus intends to evaluate the entropy of compact Vlasov objects in general relativity. In this regard, a group of scientists consisting of Naresh Dadhich, H\r{a}kan Andr\'{e}asson, Dawood Kothawala and Sahil Saini has been formed to approach this question from both theoretical and simulation route. There exists a well-established body of literature on the entropy of gravitating systems, which provides several approaches and tools to help in this task. An interesting aspect which emerged during discussions is that, while we are calculating the entropy of these systems, we could also check if the Bekenstein bound on entropy is violated. Thus, this is also included in the project. The work on this project has started.

\subsection{Possibility of detecting Buchdahl stars using gravitational waves}
Another interesting idea that emerged during the workshop was to explore the possibility of inferring the existence of Buchdahl stars from the data of gravitational waves detected from binary mergers of compact objects. As Shaswath Kapadia showed in his presentation, the gravitational wave signal contains the imprints of the tidal deformability of the detected objects. Thus, it seemed plausible that by using well-known universal model-independent relations between tidal deformability and compactness, which are routinely used in gravitational wave analysis, the detected tidal deformability can be related to the compactness of the objects. From the data of the gravitational waves detected so far, stellar remnants of a compactness close to the Buchdahl limit have not yet been observed. However, we can use these relations between tidal deformability and compactness to propose the detector sensitivity required to observe and distinguish stellar remnants which have compactness close to the Buchdahl limit but are clearly distinct from black holes. 
Discussions led to an identification of possible candidates for forming Buchdahl stars. One of the possible candidates are highly compact neutron stars or the theoretically proposed Quark stars. The other is the compact objects constituted from Einstein-Vlasov matter. It was proposed to consider both of these for this study. Thus, this particular study has two threads – one considering highly compact neutron stars or quark stars, and the other considering Einstein-Vlasov matter. A group consisting of Naresh Dadhich, H\r{a}kan Andr\'{e}asson, Shaswath Kapadia and Sahil Saini has been formed to address this question. The work on this project has also started.

\subsection{Stability of Buchdahl Stars}
A mere existence of an equilibrium state of maximum compactness is not enough to suggest that such objects would actually form in nature. The question of stability of such an equilibrium state is an important one in this respect. During the workshop, investigations approaching the question from widely different angles seemed to converge on the limiting compactness of a stable stellar remnant. H\r{a}kan Andr\'{e}asson reported in his talk that if rotation is not taken into account, his simulations of compact spherical Vlasov systems tended to become unstable at a value lower than the Buchdahl bound. A fresh search would be required to zero in on the exact value, but a rough estimate seems to be a compactness of around 0.3 (compared to the value 0.44 for Buchdahl star and 0.5 for black hole). The work by Ranjan Sharma, as reported by him in his talk, seemed to corroborate this. His work approached this question from a theoretical standpoint. He applied a method originally developed by S. Chandrasekhar to study the stability of stars under radial oscillations. When coupled with a causality criterion, Ranjan found through numerical solutions that the limiting compactness of a stable object was 0.33, beyond which instability sets in. Another piece of corroborating evidence came from the observational side: Debarati Chatterjee mentioned in his talk that the highest compactness of a neutron star observed so far was close to the 0.3 mark. The workshop has been instrumental in bringing these points of view together in one place. Thus, a project is envisaged to tie up these loose ends to come up with a limiting compactness for a stable stellar remnant. A group consisting of Naresh Dadhich, H\r{a}kan Andr\'{e}asson, Ranjan Sharma and Sahil Saini took it up. On the one side, simulations of the Einstein-Vlasov system need to be carried out to obtain a precise value of compactness beyond which the object becomes unstable. Ranjan’s theoretical analysis only checked the first order mode of perturbations. It is decided to carry out this theoretical analysis for a few higher modes of radial oscillations to ensure the credibility of the compactness bound obtained by this method. The work on this project has also started.

\subsection{Limiting speed of infalling particles to Buchdahl stars}
As argued above on theoretical grounds, a freely falling particle towards the surface of a Buchdahl star is expected to achieve very high speeds of the order of 8/9 of the speed of light. Thus, it was proposed to use H\r{a}kan Andr\'{e}asson’s simulations of compact objects, to verify whether the particle speeds approach 8/9 as the compactness is increased. This is to be taken up parallelly with the project on determining the entropy of objects of limiting compactness discussed above.

\subsection{Lifetimes of radiating Buchdahl stars}
Rituparno Goswami’s work in collaboration with Naresh Dadhich on Buchdahl stars investigated the question of radiating Buchdahl stars. Since Buchdahl stars are not black holes, they must radiate some energy to their surroundings. In their work, Rituparno Goswami, Naresh Dadhich and others found that a Buchdahl star can maintain its Virial equilibrium if the outgoing radiation is of a specific type. An interesting question pertains comparing Buchdahl stars with black holes in this respect. A well-known calculation by Stephen Hawking shows that black holes radiate energy in form of a thermal Hawking radiation and the time period of the black holes under such loss of energy has been estimated. It will be interesting to find out how similar Buchdahl stars are to black holes when it comes to this aspect. In particular, what is the rate of loss of energy for radiating Buchdahl star which maintains its Virial equilibrium, and the corresponding lifetimes can be compared to black holes of similar masses. This project will be taken up in due time, after the work on the above projects matures.

\section{Acknowledgements}
Authors would like to thank Naresh dadhich for organizing the workshop at IUCAA, bringing together a variety of scholars, and for numerous discussions.


\begin{thebibliography}{99}

     \bibitem{Naresh} Dadhich, N. (2022). On the equilibrium of the Buchdahl star. arXiv:2212.06745.
     
     \bibitem{Haken1} Andr\'easson, H. (2008). Sharp bounds on $2m/r$ of general spherically symmetric static objects. J. Differential Equations 245, 2243.

    \bibitem{Haken2} Andr\'easson, H. (2009). Sharp bounds on the critical stability radius for relativistic charged spheres. Commun. Math. Phys. 288, 715.

    \bibitem{buchdahl1959general} Buchdahl, H. A. (1959). General relativistic fluid spheres. Physical Review, 116(4), 1027.

    \bibitem{Tolman} Tolman, R. C. (1939). Static solutions of Einstein's field equations for spheres of fluid. Physical Review, 55(4), 364.

    \bibitem{Raghoonundun} Raghoonundun, A. M. \& Hobill, D. W. (2015). Possible physical realizations of the Tolman VII solution. Physical Review D, 92(12), 124005.

    \bibitem{Jiang} Jiang, N. \& Yagi, K. (2019). Improved analytic modeling of neutron star interiors. Physical Review D, 99(12), 124029.

    \bibitem{Chandrasekhar} Chandrasekhar, S. (1964). Dynamical instability of gaseous masses approaching the Schwarzschild limit in general relativity. Physical Review Letters, 12(4), 114.

    \bibitem{Chandrasekhar2} Chandrasekhar, S. (1964). The dynamical instability of gaseous masses approaching the Scwharzschild limit in general relativity, Astrophys. J., 140, 417–433.

    \bibitem{Harko2011} Harko, T., Lobo, F. S., Nojiri, S. I. \& Odintsov, S. D. (2011). $f(R,T)$ gravity. Physical Review D, 84(2), 024020.

    \bibitem{Giuliani} Giuliani, A. \& Rothman, T. (2008). Absolute stability limit for relativistic charged spheres. General Relativity and Gravitation, 40, 1427-1447.

    \bibitem{Sharma1} Sharma, R., Ghosh, A., Bhattacharya, S. \& Das, S. (2021). Anisotropic generalization of Buchdahl bound for specific stellar models. The European Physical Journal C, 81, 1-5.

    \bibitem{Saini1} Saini, S. (2018). Singularity Resolution in Anisotropic and Black Hole Spacetimes in Loop Quantum Cosmology, PhD Thesis.

    \bibitem{Boehmer} Boehmer, C. G. \& Vandersloot, K. (2007). Loop quantum dynamics of the Schwarzschild interior. Physical Review D, 76(10), 104030.

    \bibitem{Corichi} Corichi, A. \& Singh, P. (2016). Loop quantization of the Schwarzschild interior revisited. Classical and Quantum Gravity, 33(5), 055006.

    \bibitem{Saini2} Saini, S. \& Singh, P. (2016). Geodesic completeness and the lack of strong singularities in effective loop quantum Kantowski–Sachs spacetime. Classical and Quantum Gravity, 33(24), 245019.

    \bibitem{Saini3} Olmedo, J., Saini, S. \& Singh, P. (2017). From black holes to white holes: a quantum gravitational, symmetric bounce. Classical and Quantum Gravity, 34(22), 225011.

    \bibitem{AOS1} Ashtekar, A., Olmedo, J. \& Singh, P. (2018). Quantum transfiguration of Kruskal black holes. Physical review letters, 121(24), 241301.




\end{thebibliography}
\end{document}